# Mechanisms of isotope exchange between aqueous solutions and barite in low-temperature geochemical systems

Chen Zhu[1*], Youxue Zhang[2], Donald J. DePaolo[3], Kaiyun Chen[4], Honglin Yuan[4], Tao Yang[5], Lei Gong[1]

[1]Department of Earth and Atmospheric Sciences, Indiana University; Bloomington, 47405, USA.

[2]Department of Earth Sciences, University of Michigan; Ann Arbor, 48109, USA.

[3]Department of Earth and Planetary Sciences, University of California – Berkeley; California, 94720, USA.

[4]Northwest University, State Key Laboratory of Continental Evolution and Early Life; Xian 710069, China.

[5]Nanjing University, State Key Laboratory of Critical Earth Material Cycling and Mineral Deposits; Nanjing 210023, China.

*corresponding author: chenzhu@iu.edu.

**Abstract:** The prevailing view that solid-state diffusion is negligible at low temperatures is challenged by rapid sulfur and barium isotope exchange between natural barite crystals and aqueous solutions in laboratory experiments. This assumption relies on diffusivities extrapolated from high-temperature experiments. Here, isotope exchange rates were measured in solutions enriched with $^{137}$Ba and $^{32}$S at 50 and 80 °C for ≤10,360 hours. SIMS depth profiles revealed $^{137}$Ba enrichment to ~75 nm, with shapes characteristic of classical diffusion. Isotope disequilibrium between the barite surface and aqueous solution implies a continuous supply from the interior. These results indicate that defects and vacancies in barite and similar low-temperature minerals enable effective solid-state diffusion, with profound implications for paleoenvironmental reconstructions, materials science, and engineering.

# Introduction

Isotopes and trace elements of sulfate and carbonate minerals have long been used for reconstructing Earth's history and paleoenvironments. Paleo-proxies have provided records of temperatures, seawater chemistry, oxidation states, and paleo-ocean productivity (*1-4*). The fidelity of these paleo-proxies relies on the assumption that these minerals have preserved their original isotopic and trace element signatures over millions of years after their initial deposition and have escaped diagenetic overprinting by post-formational processes. However, a growing literature in recent years indicates that isotopic exchange and recrystallization occur rapidly in laboratory settings, and if extrapolated, suggests that the isotopes in these proxies can be altered at a time scale of decades to thousands of years, much shorter than the millions of years that these proxies are supposed to represent (*5-14*).

Evaluation of the fidelity of paleo-proxies requires both innovative experimental designs that can measure reaction rates at solubility equilibrium and resolving the reaction mechanism conundrum with the state-of-the-art analytical tools. The deep geological time scale in studying Earth's history posits that sulfate and carbonate minerals must have reached solubility equilibrium with pore fluids, a process that only takes days in the laboratory. The relevant chemical conditions are, therefore, near chemical or solubility equilibrium rather than far from equilibrium, where most kinetic data are currently available. However, isotope tracers are known to illuminate chemical reactions at equilibrium (*15-18*) and have been applied recently with Ba isotope doping to derive isotope exchange rates with barite at solubility equilibrium (*19*). Isotopic tracers make studying the reactivity of paleo proxy minerals at solubility equilibrium possible.

Additionally, extrapolating short-term laboratory experimental rates (e.g., $<10^0$-$10^1$ years) to geological time scales ($>10^3$-$10^6$ years) is plausible only if we understand the underlying reaction mechanisms that are associated with the laboratory-measured rates. Laboratory-based far-from-equilibrium reaction rates are $10^2$–$10^5$ times faster than estimated field rates (e.g., *20, 21*). Therefore, we must ascertain that the mechanisms of reactions operating in the laboratory and geological systems are comparable.

In this study, we carried out four series of barite-aqueous solution isotope exchange batch experiments. The starting solutions of each series were doped with $^{137}$Ba and $^{32}$S and were drastically different in these isotopes from the natural barite reactant. Reacted barite and solution samples were collected for isotope analysis after 504, 1008, 1512, 8208, and 10,360 hours of reactions at 50 and 80 °C at pH ~4.5. SIMS depth profiles of $^{137}$Ba in reacted barite grains in selected experiments were used to determine isotopic penetration from the grain surfaces, and the temporal evolution of S and Ba isotopes in reacted solutions provided the overall barite-water exchange rates. Modeling was carried out to interpret the SIMS depth profiles of $^{137}$Ba and associated aqueous isotope data. Barite was chosen for this study because it has long been used as a model mineral for studies of nucleation and growth/dissolution rates of sparingly soluble crystalline compounds (*22*). The trace elements and isotopes in marine barite have provided critical information on the history of seawater chemistry and ocean paleo-productivity (*1-3*). Understanding the isotope exchange rates for barite has significant implications for various geochemical paleo-proxies, such as those involving carbonates.

## Results

$^{137}$Ba was introduced into the starting solutions, which resulted in significantly higher $^{137}$Ba fractions ($f_{137}$ of 0.5555 in 50-1 and 80-1 series and 0.2989 in 50-2 and 80-2 series, where 50 and 80 indicate the experimental temperature) than the natural barite reactant ($f_{137}$ of 0.11232). The starting solution also had a $\delta^{34}$S (-2.9‰ VCDT) composition significantly lower than the barite crystals ($\delta^{34}$S +22.7‰ VCDT). SIMS $^{137}$Ba/$^{138}$Ba depth profiles from cleavage surfaces of reacted grains indicate that isotope exchange reached ~75 nm deep into barite grains after 10,360 hours of reactions. The reacted barite shows a 62% increase in $^{137}$Ba/$^{138}$Ba ratios at the barite-water interface (**Fig. 1**) while the unreacted barite control showed constant $^{137}$Ba/$^{138}$Ba ratios (±0.3% variation; 1σ) from the grain surface to the interior (blue line). At a depth of ~75 nm into the grain, $^{137}$Ba/$^{138}$Ba matches the unreacted barite composition, indicating that the underlying $^{137}$Ba/$^{138}$Ba of the unreacted part has been reached. The authenticity of $^{137}$Ba/$^{138}$Ba enrichment was further verified by the almost constant ratios of undoped $^{136}$Ba/$^{138}$Ba ratios (as internal control) from the surface to the interior (orange line). Time series samples showed progressive intrusion of $^{137}$Ba into the grain interior with time. In **Fig. 1**, sample 80-1-4 reacted for 8,208 hours while sample 80-1-5 reacted for 10,360 hours. Attempts to profile $^{34}$S/$^{32}$S were not successful due to insufficient ion probe signals.

One barite monolayer is about 0.351 nm (*23*). A ~75 nm reacted thickness suggests that ~214 monolayers have participated in the Ba isotope exchange reactions, which is far beyond the "distorted, relaxed, and high ion-mobility" first four monolayers or two unit-cells on barite surface (*23*). In short, the SIMS profiles exhibit smooth, diffusion-like gradients rather than sharp fronts, extending up to ~75 nm. These profiles are consistent with classical diffusion (*24*) and extend well beyond the expected surficial reaction layers.

Cross-sections of Ba isotopes were also measured using Laser Ablation MC-ICP-MS (LA) with a spot size of ~ 7 μm (**Supplementary Materials (SM), S9**). Because of the large spot size, these data could not resolve the short profiles near the surface, but provided a broader spatial overview. First, the 120 cross-sections showed that $^{137}$Ba/$^{134}$Ba at the grain edges increased by 12 to 74% compared to the grain interiors (**fig. S4a**). The observed $^{137}$Ba enrichments were supported by nearly constant $^{136}$Ba/$^{134}$Ba ratios at the same locations. Thus, the LA data provided robust statistical validation that the deep isotopic penetration observed by SIMS is a widespread process. Second, the $^{137}$Ba/$^{134}$Ba ratios in the interior of the unreacted barite grains are homogeneously distributed at the LA beam resolution. Third, five out of 120 cross-sections showed $^{137}$Ba/$^{134}$Ba peaks in the interior of the barite grain, while $^{136}$Ba/$^{134}$Ba are constant (serving as internal standard), which indicates $^{137}$Ba isotope exchange along barite cleavages or cracks (**fig. S4b**).

*Ba and S isotope exchange rates from the aqueous solution data*

At the end of the experiments, 0.25 to 0.3 fractions of Ba and S isotopes in the fluids had exchanged with barite (**Fig. 2**). The fraction of $^{137}$Ba, $f_{137}$, experienced a decrease of ~0.1 and $\delta^{34}$S increased by ~7‰ (**Fig. 3**). The temporal evolution of Ba and S isotopes was used to calculate the Ba and S isotope exchange rates (**SM, S7**) using the integrated first-order equation (*25*),

$$\ln(1-F) = -kt \qquad (1)$$

where $k$ is the bulk rate constant in s$^{-1}$, $t$ is the time in s, and $F$ is the fraction of the isotopes in fluids that have exchanged with barite (**Eq. S5**).

The experimental data fitted well into the first-order equation (**Eq. 1**), but it appears that exchange rates were faster before 504 and 1008 h and then maintained a slower but steady-state rate until the end of the experiments (10,360 h). Such a transition and decrease within a few hundred hours of reactions were also seen in calcite isotope exchange experiments (*7-9*). Therefore, we used **Eq. (1)** to calculate exchange rates at two time-segments, 0 to 1008 h and 1008 to 10,360 h. The fast reaction period yields a bulk exchange rate of 8.43 ×10$^{-10}$ mol s$^{-1}$ m$^{-2}$ at 50 °C for Ba. For the long-term exchange rates, we derived 4.84 × 10$^{-11}$ mol s$^{-1}$ m$^{-2}$ at 80 °C and 2.11 × 10$^{-11}$ mol s$^{-1}$ m$^{-2}$ at 50 °C for both Ba and S isotopes, yielding an apparent activation energy ($E$a) of 26.2 kJ mol$^{-1}$. This value compares with 30.8 kJ mol$^{-1}$ measured by Dove and Czank (*26*) for far-from-equilibrium dissolution rate constants.

The mass balance considerations of the isotopes in the aqueous solutions require a continuous background isotope flux from the solid interior toward the surface (*27*), which is macroscopic evidence corroborating diffusion-like profiles observed by SIMS and LA-MC-ICP-MS. This continuous isotope flux demonstrates that the process is not merely a localized surface effect on a few grains but contributes to the overall system isotopic evolution. Furthermore, at the end of the experiments (10,360 h), the barite surfaces and aqueous solutions did not reach isotope equilibrium, and isotope exchange continued.

## Discussion

### *Modeling Ba isotope profiles in barite*

The smooth profiles of $^{137}$Ba fraction ($f_{137}$) spanning 20 to 50 nm are consistent with diffusion-controlled transport. Infiltration along cleavages and/or cracks would not produce such smooth profiles. Nonetheless, cleavages and/or cracks played a role in larger-scale variations (micrometer scale) (**fig. S4b**).

The general case of Ba isotope exchange between aqueous solution and barite includes both boundary motion (dissolution or growth of barite) and diffusion of Ba isotopes into a chemically homogeneous barite. Hence, the advection-diffusion equation in the dissolving crystal (e.g., Zhang, 2008, eq. 4-102a) (*24*) is as follows:

$$\frac{\partial f}{\partial t} = \frac{\partial}{\partial y}\left(D \frac{\partial f}{\partial y}\right) + U \frac{\partial f}{\partial y}, \quad y > 0, \ t > 0 \tag{2}$$

where $y$ denotes the distance in barite measured from the barite surface in an interface-fixed reference frame, and $f$ is the fraction of a given isotope. Because Ba concentration in barite ($C_{Ba}^{barite}$) is constant (19143 mol m$^{-3}$), the isotope fraction serves as concentration in the diffusion equation. $U$ is the barite dissolution rate (m s$^{-1}$), and $D$ is the diffusivity of the barite isotope (pm$^2$ s$^{-1}$). Because diffusion profiles are short (< 0.1 μm) and barite particles are large (> 100 μm), the diffusion problem can be treated as one-dimensional semi-infinite space.

The boundary condition at the barite surface ($y = 0$) needs to be carefully derived. DePaolo and Zhang (*27*) derived the following boundary condition:

$$D \left(\frac{\partial f}{\partial y}\right)_{y=0} = \frac{u_d}{C_{Ba}^{barite}}(f_{bs} - f_{aq}) - \frac{u}{C_{Ba}^{barite}} f_{bs}, \tag{3}$$

where $u_d$ is the detachment rate (also referred to as the backward reaction rate $R_b$) in mol m$^{-2}$ s$^{-1}$, $u = u_d - u_a$ is net dissolution rate (also referred to as net reaction rate $R_{net}$), $u_a$ is the attachment rate (also referred to as the forward reaction rate $R_f$), and subscript bs and aq in $f$ means barite surface and aqueous solution, respectively. The above boundary condition applies when $u = 0$. If $u \neq 0$, the above boundary condition would lead to an erroneous non-zero $\left(\frac{\partial f}{\partial y}\right)_{y=0}$ even if $f_{bs} = f_{aq}$. However, if barite has the same $f$ as the aqueous solution, there should not be a variation of $f$ in barite unless diffusive isotope fractionation is considered. Hence, the boundary condition is re-derived (**SM, S10**), leading to:

$$D \left(\frac{\partial f}{\partial y}\right)_{y=0} = \frac{w u_d}{C_{Ba}^{barite}} (f_{bs} - f_{aq}), \tag{4}$$

where $w = [Ba^{2+}][SO_4^{2-}]/K_{sp}$ is the degree of saturation of the aqueous solution.

Define $U_d = u_d/C_{Ba}^{barite}$. $U_d$ has the unit of m s$^{-1}$ and is the linear detachment rate. The dissolution rate $U$ in **Eq. (2)** can be related to $U_d$ and $w$ as $U = U_d(1-w)$ (*24, 27*). Note that the $f_{aq}$ in the above equation depends on time (**SM, S10**). Given the diffusion **Eq. (2)** and the boundary condition in **Eq. (4)**, the $f_{137}$ profiles in barite were fitted with two fitting parameters, $D$ and $u_d$. The fits of $^{137}$Ba fraction ($f_{137}$) profiles in barite and calculated $f_{137}$ at barite surface as a function of time are shown in **Fig. 3** and the fitting parameters are presented in **Table 1** and **table S6**. All fits yielded high-quality regression, with minor systematic deviations likely attributed to a second, more rapid diffusion mechanism (*28*), which affects a small fraction of Ba atoms. The use of a $u_d$ in the 1008 hours 40 times that in the subsequent hours does not noticeably change the fit quality of $^{137}$Ba profile in barite.

The calculated evolution of $f_{137}$ on barite surface is displayed in **Fig. 3c,d**. Also shown are the measured $f_{137}$ in the aqueous solution as a function of time and the measured $f_{137}$ on barite surface for 80-1-4, 80-1-5a,b, and 80-2-5. The SIMS-measured $f_{137}$ on barite surface for 80-1-4 falls on the calculated barite surface curve for 80-1-5. The black dashed curve in **Fig. 3d** is calculated if $u_d$ is increased 10 times: with increasing $u_d$, the surface $f_{137}$ on barite would track the aqueous solution data after a short transient period, which does not agree with $f_{137}$ on the barite surface (purple square).

The fitted detachment rate $u_d$ of Ba from barite surface to aqueous solution at 80° C ranges from $1.5 \times 10^{-12}$ to $1.9 \times 10^{-11}$ mol m$^{-2}$ s$^{-1}$ ($1.7 \times 10^{-12}$ to $2.2 \times 1^{-11}$ mol m$^{-2}$ s$^{-1}$ if minor dissolution is allowed at the beginning of the experiments). The values of $u_d$ from our experiments are dictated by $(f_{bs} - f_{aq})$. Increasing $u_d$ would lead to $f_{bs}$ values tracking $f_{aq}$ values after a short transition time (e.g., black dashed curve in **Fig. 3d**), not supported by experimental data.

The obtained self-diffusivity of Ba in barite at 80° C ranges from 1.3 to 8.3 pm$^2$ s$^{-1}$ (1 pm$^2$ s$^{-1}$ = $1 \times 10^{-24}$ m$^2$ s$^{-1}$). The variation of $u_d$ and $D$ in the 80° C experiments might be attributed to: (1) anisotropy, and (2) inconsistency in types and concentrations of defects. A factor of six scatter for diffusivity in a crystalline phase is not surprising. For example, even at high temperatures at which the defect concentrations are expected to be more equilibrated, $^{13}$C and cation diffusivities in calcite (*29*) and cation diffusivities in olivine (*29*) can still vary by orders of magnitude depending on the experimental and analytical methods.

The co-variation of S and Ba isotope ratios implies that S self-diffusivity in barite is about the same or slightly greater than Ba self-diffusivity (**SM, S10**).

*Comparison with literature data*

There are no direct data for Ba diffusion in barite at any temperature for comparison. Our obtained Ba self-diffusivity values in barite are ~13 to 42 orders of magnitude greater than those expected from Arrhenius extrapolation of $^{44}$Ca diffusion coefficients measured over the temperature range 550 - 900 ºC on annealed calcite (*30*). For example, extrapolated $^{13}$C, $^{42}$Ca, or $^{44}$Ca self-diffusivities to 80° C in calcite range from $10^{-42}$ to $10^{-13}$ pm$^2$ s$^{-1}$ (*29*). Hence, it's clear that diffusivities at low temperatures cannot be extrapolated from high-temperature data. Some argue that minerals at low temperatures may have both nano- or micro-porosity, and more defects and vacancies so that aqueous-mediated faster diffusion is possible and different from bulk lattice diffusion (*7, 27, 31*).

For calcite, estimates of diffusivities at low temperatures are available in the literature. Evidence for near calcite surface diffusion of $^{45}$Ca and $Cd^{2+}$ at ambient temperatures was presented in 1960s and 1990s (*32, 33*). Recently, Géhin et al. (*7*) estimated $^{13}$C self-diffusivity from bulk $^{13}$C uptake data in synthetic calcite over time to be 0.48 pm$^2$ s$^{-1}$ at 21° C and 1.5 pm$^2$ s$^{-1}$ at 50° C. In other words, observed sustained $^{13}$C uptake by calcite is beyond what could be explained by a single monolayer exchange. From mass balance, they estimated that about eight layers or a depth of 3.8 nm are involved after one month of reaction, using a single monolayer of calcite thickness of 0.5 nm.

Harrison et al. (*9*) exposed natural and synthetic calcite to fluids highly enriched in $^{13}$C and $^{18}$O and measured the average isotopic composition of the solid calcite as a function of time. They estimated diffusion coefficients for $^{13}$C (ranging from 0.042 to 11 pm² s$^{-1}$) and for $^{18}$O (ranging from 0.02 to 10 pm² s$^{-1}$), which were derived by employing a "homogeneous model".

Effective diffusion constants for synthetic and natural calcite were also estimated from temporal aqueous solution data. DePaolo and Zhang (*27*) developed an advection-diffusion-reaction model and fitted the detachment rates from Naviaux et al. (*34*) to obtain a diffusivity of 3 pm$^2$ s$^{-1}$ at 5 °C for a synthetic $^{13}$C-labeled calcite. Harrison et al. (*8*) reacted synthetic nano-sized calcite and aragonite in $^{43}$Ca-enriched solutions and analyzed δ$^{43}$Ca in solutions and average values in solids. From the aqueous solution data, they estimated diffusivities of 1 pm$^2$ s$^{-1}$ at 25 °C for calcite, using the "homogeneous model," which assumes that the recrystallized fraction of the solid is in isotopic equilibrium with the fluid phase at all times.

In general, like what we found in this study for barite, the continued evolution of aqueous solution isotope compositions suggests that surface layers on the solids do not reach isotopic equilibration with the fluid. Rather, mass balance considerations require a flux from the solid interior to the solid-water interface, which continuously supplies isotopes to the surface layer (*27*). As pointed out by Gorski and Fantle (*5*) and Heberling et al. (*31*), the solid fraction cannot be unambiguously estimated from the mass balance using the solution isotope composition alone.

SIMS depth profiles provide direct evidence of diffusion-like processes that contrast with the indirect evidence of macroscopic isotope uptake in the solids or aqueous solution evolution discussed above, but high-quality SIMS data are challenging to procure. Previously, Subhas et al. (*35*) reported $^{13}$C diffusion profiles in calcite, which were analyzed by DePaolo and Zhang (*27*) to obtain C self-diffusivity of about 10 pm$^2$ s$^{-1}$ in calcite at 21°C. However, Subhas et al. (*35*) used a synthetic calcite crystal and reacted the calcite via a polished surface, which could have damaged the surface and resulted in a higher *D* than a natural cleavage surface. Their reaction time of 48 hours was short, and the penetration depth of ~2 nm, pertaining to about four monolayers, which still involves the relaxed and distorted surficial layers (*23*) but not extending significantly into the bulk crystal.

Unlike previous studies, which primarily relied on bulk uptake or short-term experiments, this study combines high-resolution SIMS profiles with LA-MC-ICP-MS data and solution chemistry to provide a multi-scale view of isotope exchange mechanisms. The robust, high-resolution SIMS profiles provide the direct visual proof. These are then broadly corroborated by the more numerous, albeit qualitative, LA-MC-ICP-MS cross-sections, which confirm widespread isotopic penetration. Finally, the macroscopic aqueous solution isotope data provide independent evidence of continuous, deep exchange, with calculated rates that align with the microscopic observations. This integration of different scales and methods of observation strongly supports the conclusion that a solid-state diffusion-like process is a significant factor in isotopic alteration.

*Other hypotheses of isotope exchange reaction mechanisms*

In addition to solid-state diffusion, there are other hypotheses that have been proposed to explain exchange reactions extending beyond surface layers, mostly for interpreting aqueous solution (macroscopic) data (*5, 27, 31, 36*). These include Ostwald ripening (dissolution of small grains and growth of larger grains) (*19, 21, 25*), interfacial dissolution-reprecipitation (*26-29*), or dissolution on some crystal faces and growth on other faces (*37, 38*).

DePaolo and Zhang (*27*) explained that the Ostwald effects are not quantitatively large enough to explain the observation of extended isotope exchange. Géhin et al. (*7*) stated that Ostwald ripening plays a small or negligible role in their model interpretation of experimental results. The "interfacial dissolution-reprecipitation" model has been proposed to explain a wide variety of silica glasses, crystalline silicates, carbonates, and sulfates (*39-42*). The original mineral phase is dissolved, and a new mineral phase is simultaneously precipitated. The dissolution creates a porous interface between the original mineral and the fluid. Additionally, pre-existing microfractures, defects, or grain boundaries within the mineral can serve as conduits for fluid ingress, enhancing the mineral's susceptibility to alteration. Such a mechanism often produces a sharp transition in SIMS profiles (*39, 43*), which is in stark contrast with our SIMS profiles.

While the above discussion is all about isotope exchange, these discussions should also be applicable to trace elements, which are also widely used for paleoenvironmental proxies. There are similarities between the microscopic models for isotope and trace element incorporation and partitioning (*36, 44, 45*). The results of this isotope study will also help test the fidelity of trace element-based paleo-proxies.

Additionally, the SIMS depth profiles in this study resolve the debate of "homogeneous" or "heterogeneous" models for isotope exchange. The homogeneous model assumes continuous re-equilibration between the recrystallized fraction of the mineral and solution, while the heterogeneous model assumes no back reaction (*5, 31*). It is clear from our data that isotope distribution in the reacted solids follows a pattern of classic diffusion profiles, which fits neither the homogeneous model nor the heterogeneous model.

*The physical meaning of isotope exchange rates*

While the physical meaning of detachment rate, $u_d$, in **Eq. (4)** is well defined as the unidirectional detachment (dissolution) rates due to the mineral-water interface reaction (*19, 44*), the isotope exchange rate calculated from **Eq. (1)** does not carry any implications of the reaction mechanism (*25*). So, what are the isotope exchange rates in our experiments?

The isotope exchange rates are the isotope fluxes from the aqueous solution into the solid, and vice versa, from solids to the aqueous solution, recorded in the temporal evolution of the isotopic compositions of Ba isotopes, in the form of $F$. At solubility equilibrium, $u_d = u_a$. The detachment reaction changes the aqueous solution Ba isotope composition, but the attachment reaction does not, in strongly doped experiments in which isotope fractionation is negligible (*46*).

From the modeling work in this study, the temporal evolution of $f_{137}$ (**Fig. 3c,d**) is determined by both $u_d$ and $f_{bs}$. $f_{bs}$ is determined by $u_a$, $f_{aq}$, and $D$ values. Therefore, the isotope exchange here is controlled by both diffusion in barite and interface reaction (*24*). The fitted $u_d$ is weakly negatively coupled with $D$, which is consistent with the sensitivity analysis in DePaolo and Zhang (*27*). Therefore, isotope exchange rates at chemical equilibrium are not net dissolution rates or unidirectional dissolution rates, and it is not meaningful to compare directly to the far-from-equilibrium net dissolution rates.

On the other hand, $u_d$ are constant when the system moves from far from equilibrium to solubility equilibrium if reaction mechanisms and reactive surface areas remain constant. Then, $u_d$ would be equal to the net dissolution rate far from equilibrium. However, the derived $u_d$ values are four orders of magnitude slower than net dissolution rates measured at far from equilibrium (*26*). This discrepancy likely arises from the depletion of reactive sites, such as kinks and steps, which are not continuously regenerated under conditions of solubility equilibrium (*19, 45*).

## Implications

Although our study provides robust evidence for solid-state diffusion in barite, and an estimate of diffusion coefficient at 80 °C, extrapolating laboratory results to natural settings over geologic timescales is challenging. For example, our diffusion-aided exchange rates suggest that solid state diffusion alone could significantly alter barite (and calcite (*47*)) after thousands of years. This may occur in nature, but it is not consistent with the apparent preservation in calcite and barite of primary seawater records that yield coherent climate and tectonic histories across diverse locations (*1, 48*).

Several previous studies have estimated long-term diagenetic solid-fluid exchange rates for calcite in deep sea sediments and found extremely slow, but significant rates that decrease systematically roughly as sediment age$^{-1}$ (e.g., *49*). This 1/t pattern is different from the expectation for a diffusion-controlled process ($1/t^{1/2}$), which leads to the conclusion that solid state diffusion is not the primary control on fluid-solid diagenetic exchange because it is too fast (*27*). More likely, diagenetic exchange is limited by the slow mineral surface exchange with fluid (*27, 49*). Near chemical equilibrium, mineral surface reactivity is reduced sharply compared to far from equilibrium conditions and decreases with time (e.g., *50*).

Nevertheless, our findings underscore two key points for paleoenvironmental studies: (1) meticulous sample selection is essential in field studies to minimize diagenetic overprinting of primary signals, and (2) the observed rapid laboratory reactions highlight an opportunity to actively target and decipher post-depositional paleoenvironments of the most recent significant event. Ultimately, understanding the mechanisms of isotope exchange between barite (and calcite) and aqueous solutions at solubility equilibrium represents a fundamental advance with profound implications for reconstructing Earth's environmental history and for isotope and trace element geochemistry.

Our results also have implications for materials science and engineering. Many materials used in infrastructure, electronics, or other applications are exposed to low-temperature aqueous environments over long periods. If components (e.g., trace elements, isotopes) can exchange and diffuse within solid phases faster than expected, it could lead to unanticipated rates of material alteration, degradation, or changes in mechanical properties, affecting their service life and safety. For example, in studies of waste disposal, barite has been proposed as a candidate to encase radioactive and toxic metals for long-term storage. Barite's ability to incorporate Ra, Sr, and Pb offers the possibility of using it to sequester those elements in radioactive waste disposal (*31, 51, 52*), but our discovery of high effective diffusivities calls for more careful consideration of barite's long-term properties.

In general, a more accurate understanding of low-temperature diffusion mechanisms could lead to new approaches in materials synthesis, allowing for better control over the incorporation or exclusion of specific elements, or predicting the stability of new composite materials.

**Acknowledgments:** Much of the manuscript writing was completed while the senior author was visiting the University of Cambridge in 2023 and 2025. CZ acknowledges the support of a Leverhulme Visiting Professorship grant that supported his visits and expresses his gratitude to Professor Nicholas Tosca and the Department of Earth Sciences at Cambridge. Dr Yunbin Guan at Caltech is acknowledged for obtaining the SIMS profiles in this study.

**Funding:**

Haydn Murray chair endowment at Indiana University to CZ.

Leverhulme Trust visiting professorship grant to the University of Cambridge with Nicholas Tosca as the PI (to host CZ).

US NSF grant EAR-2314724 to YXZ.

U.S. Department of Energy, through its Geoscience program at LBNL under Contract DEAC02-05CH11231 to DJD.

**Author contributions:**

Conceptualization: CZ, YXZ, DJD

Methodology: CZ, YXZ, LG, YBG, HLY, TY

Investigation: LG, CZ, YXZ

Visualization: LG, YBG, YXZ, CZ

Funding acquisition: CZ

Project administration: LG

Supervision: LG

Writing – original draft: CZ, YXZ

Writing – review & editing: CZ, YXZ, DJD

**Competing interests:** Authors declare that they have no competing interests.

**Data and materials availability:** All data are available in the main text or the supplementary materials.

## Supplementary Materials

Materials and Methods

Figs. S1 to S4

Tables S1 to S6

References (53-69)

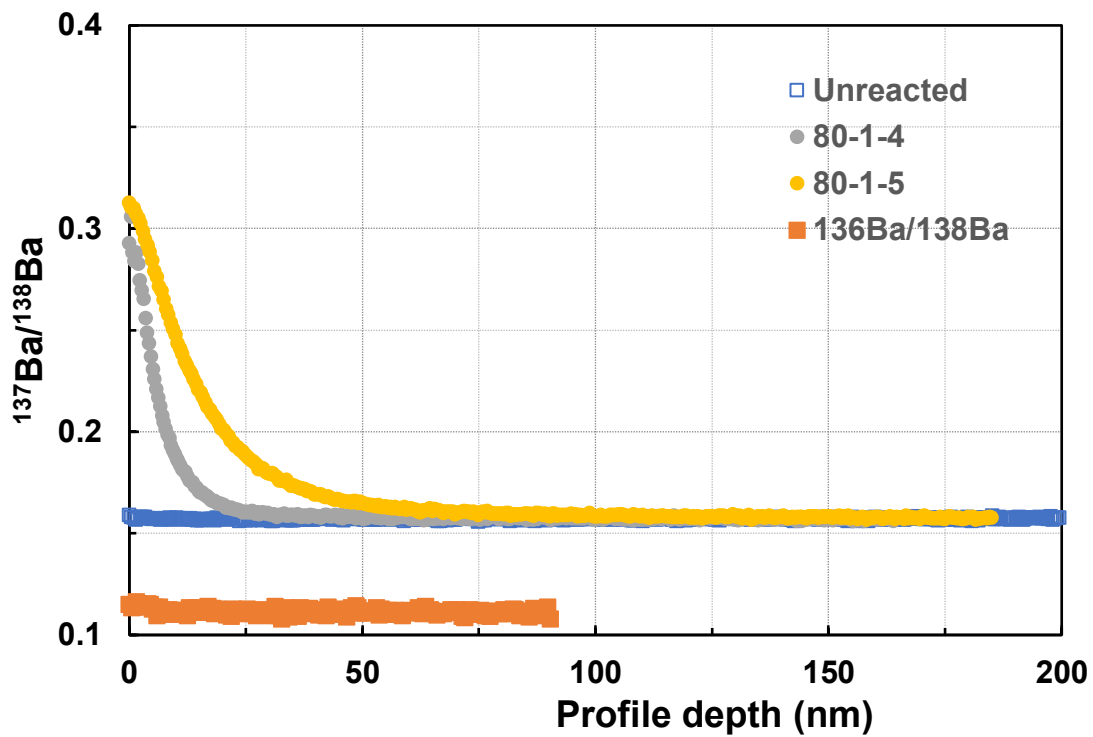

**Fig 1.** **$^{137}$Ba/$^{138}$Ba SIMS depth profiles of barite grains**. Barite grains were reacted with solutions for 8,208 (grey) to 10,236 (orange) hours, respectively, at 80 °C. The blue line is the unreacted barite control. $^{137}$Ba enrichment demonstrates that $^{137}$Ba from an aqueous solution (with a $^{137}$Ba/$^{138}$Ba of 1.48 in the starting solution) has been incorporated into the bulk crystalline barite beyond the relaxed surficial layer of two unit-cells or ~1.4 nm. The orange line shows that $^{136}$Ba/$^{138}$Ba does not change from the boundary to the core, which serves as internal control and supports the authenticity of observed $^{137}$Ba/$^{138}$Ba enrichment at the barite edge.

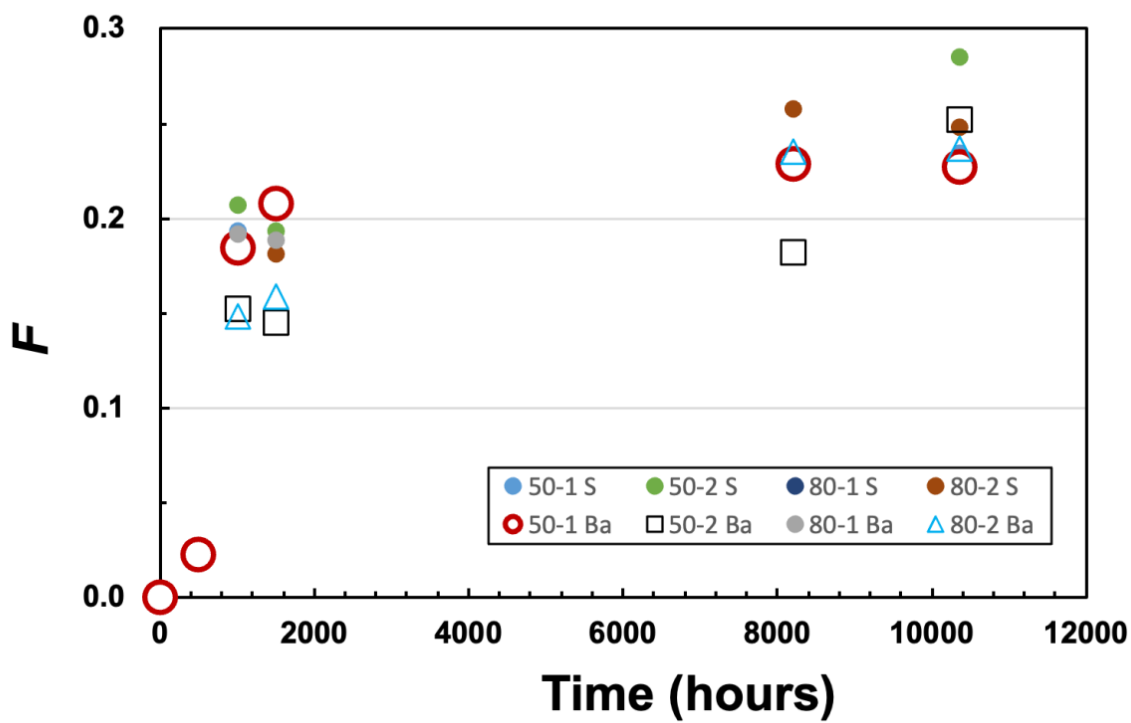

**Fig. 2. Fractions of S and Ba isotopes exchanged**. $F$ is calculated with (Eq. S5) and represents the fractions of isotopes in the aqueous solutions that have exchanged with barite. For symbol labels, "50" and "80" denote experimental temperature in °C, and "-1" and "-2" denote doping strength of $^{137}$Ba ($f_{137}$ = 0.5555 or =0.2989, respectively). Theoretically, $F$ should be zero at time zero.

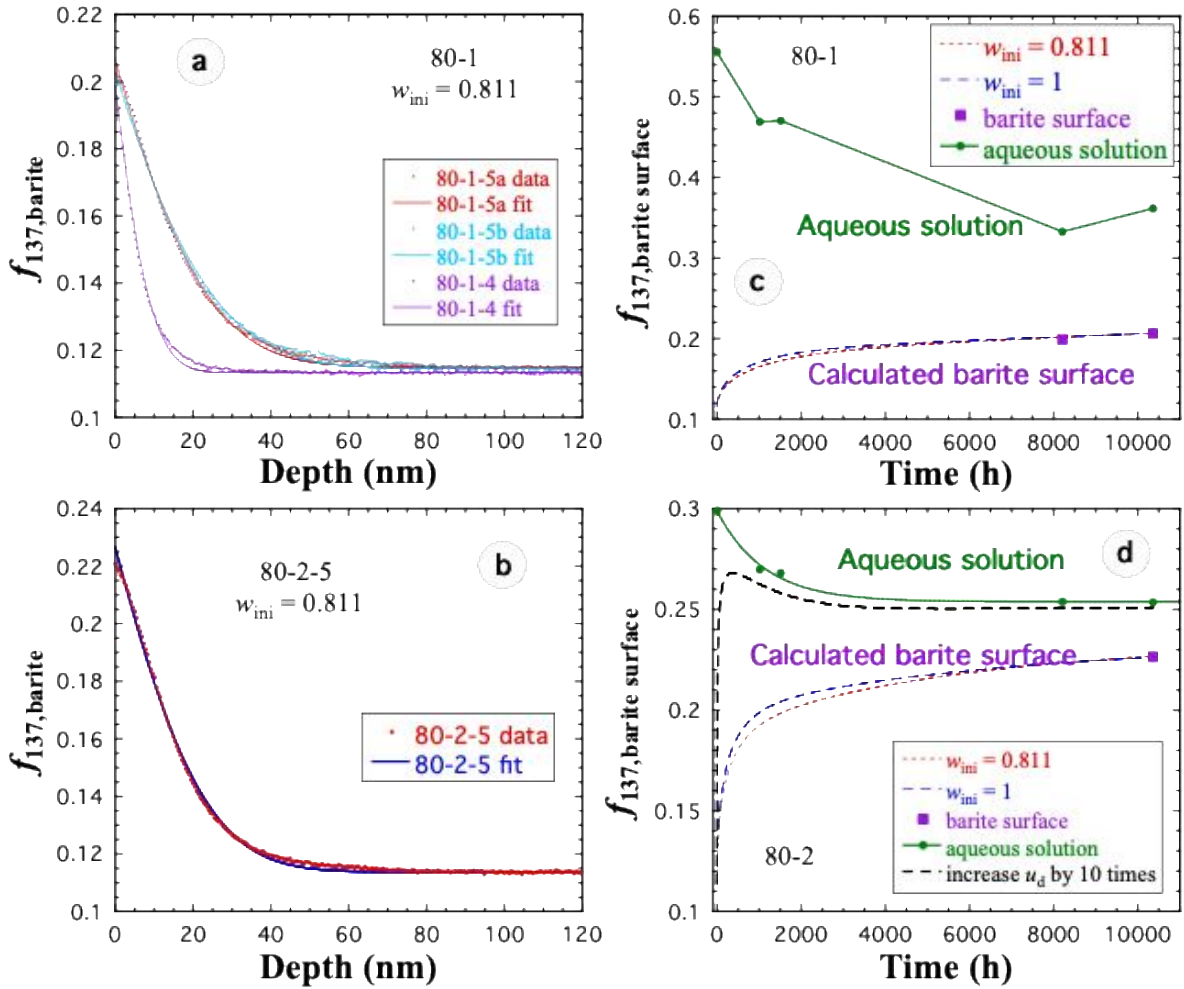

**Fig. 3. Modeled diffusion profiles in barite solids and $f_{137}$ in the aqueous solutions.** (a) and (b) Depth profiles of $f_{137}$ in barite and fit curve. (c) and (d) The calculated $f_{137}$ evolution at barite surface as a function of time based on modeling diffusion profiles in (a) and (b), compared with $f_{137}$ at barite surface (purple squares) measured with SIMS. The black dashed curve for surface $f_{137}$ in (d) was calculated by increasing $u_d$ by 10 times. The $f_{137}$ evolution in the aqueous solution is shown as green solid symbols (experimental data) and calculations.

Table 1. Fit results of $f_{137}$ diffusion profiles in barite assuming $w_0 = 1$

| Profile | Time (hr) | $D_{Ba}$ (pm$^2$s$^{-1}$) | $u_d$ (10$^{-11}$ mol m$^{-2}$ s$^{-1}$) | $f_{137,y=0}$ | $\Delta M_{137}$ (10$^{-5}$ mol m$^{-2}$) | $r^2$ |
|---|---|---|---|---|---|---|
| 80-2-5 | 10360 | 5.49±0.09 | 1.944±0.04 | 0.22623 | 3.25 | 0.99784 |
| 80-1-5a | 10360 | 6.53±0.11 | 0.370±0.003 | 0.20650 | 2.77 | 0.9974 |
| 80-1-5b | 10360 | 7.92±0.15 | 0.381±0.003 | 0.20221 | 2.91 | 0.9967 |
| 80-1-4 | 8208 | 1.24±0.03 | 0.154±0.002 | 0.19862 | 1.00 | 0.9939 |

All experiments were at 80 °C. The indicated errors are at 2σ level. 1 pm$^2$ s$^{-1}$ = 10$^{-24}$ m$^2$ s$^{-1}$. $\Delta M_{137}$ is the total (integrated with respect to time) number of moles of $^{137}$Ba that entered barite from the aqueous solution based on the diffusion profile.

# Supplementary Materials for

## Mechanisms of isotope exchange between aqueous solutions and barite in low-temperature geochemical systems

Chen Zhu[1*], Youxue Zhang[2], Donald J. DePaolo[3], Kaiyun Chen[4], Honglin Yuan[4], Tao Yang[5], Lei Gong[1]

Corresponding author: chenzhu@iu.edu.

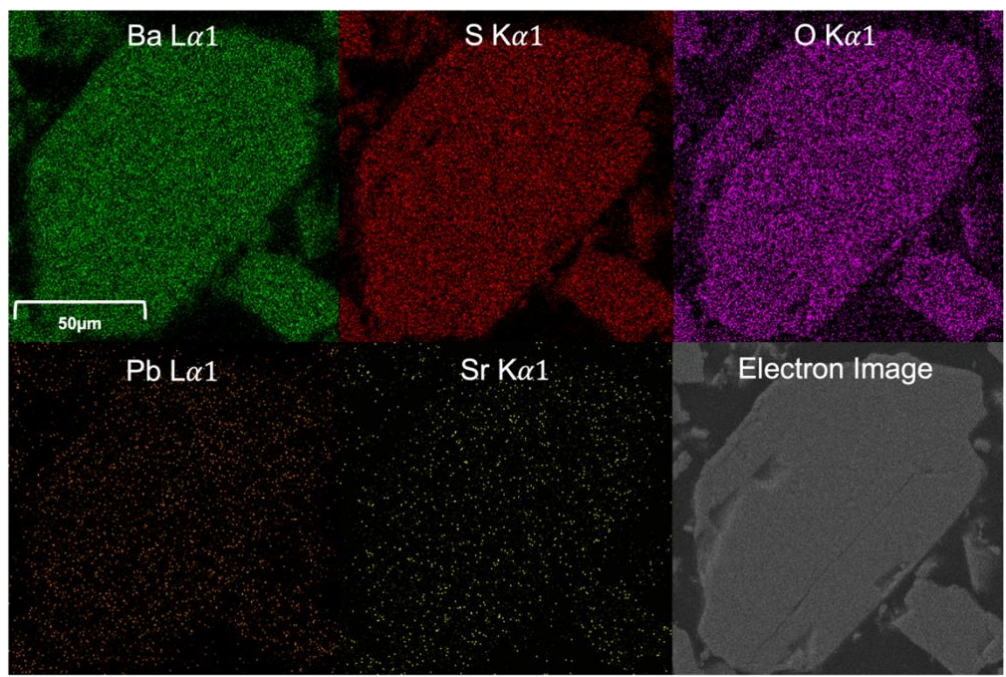

**Fig. S1. Scanning electron micrograph of barite grains.** EDX maps of one barite grain (unreacted) in a petrographic thin section, showing high-purity barite without much Pb or Sr impurity.

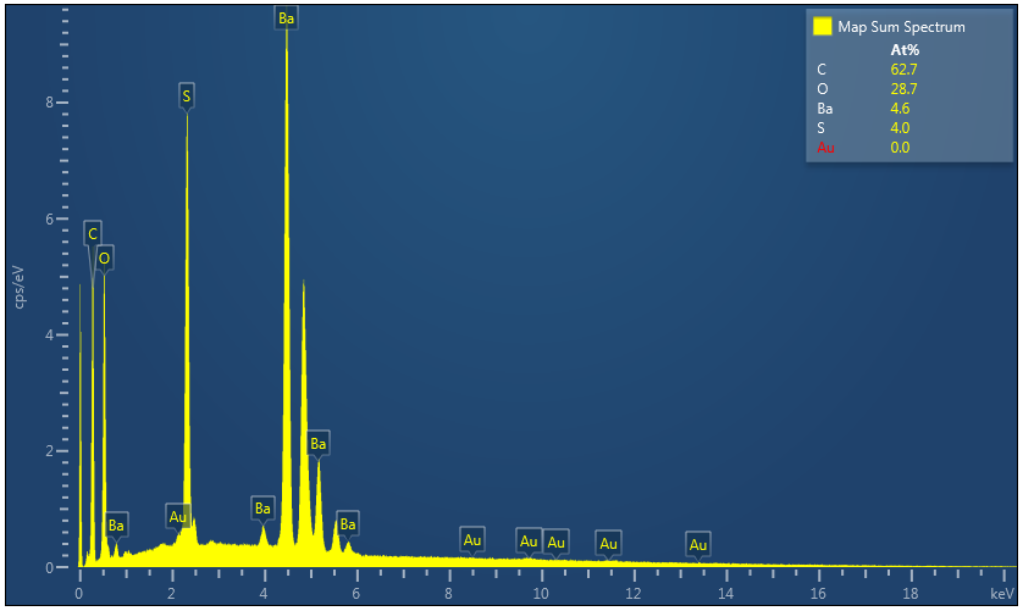

**Fig. S2. EDX spectral results of unreacted natural barite**. The spectra show the counts per second per electron volt for each element and the atomic percentage representing the proportion of each element in grain one. The carbon and gold atomic percentages relate to the matrix in the thin section material rather than the grain. Oxygen has the highest atomic percentage, followed by barium and sulfur.

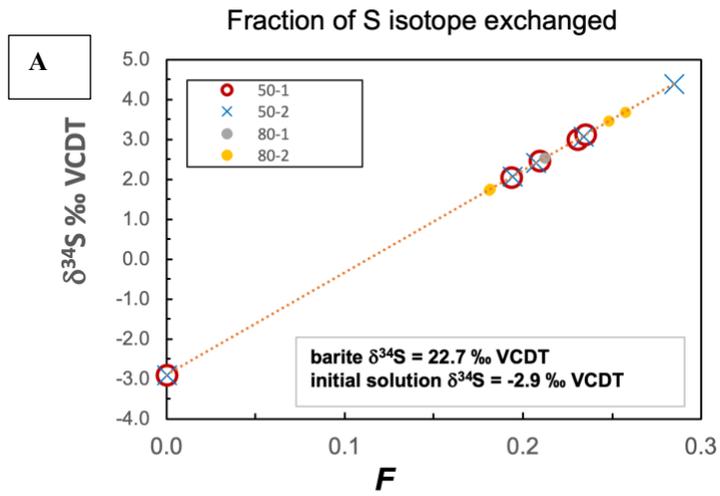

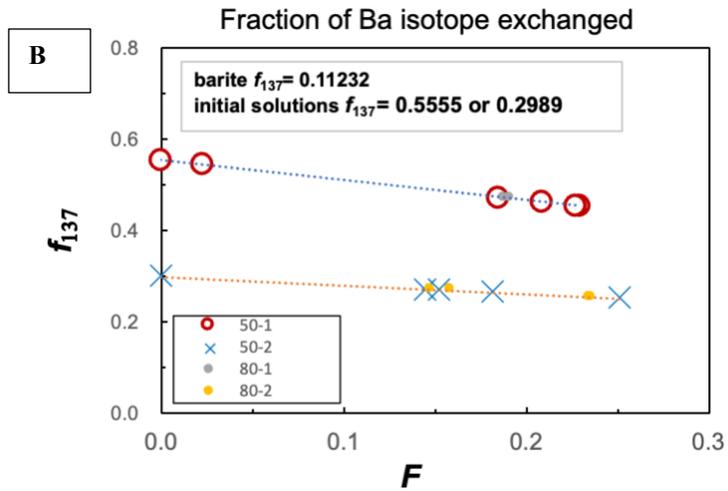

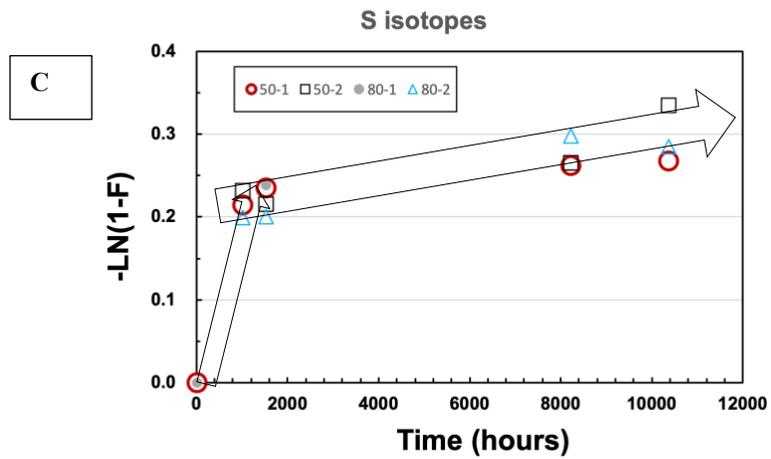

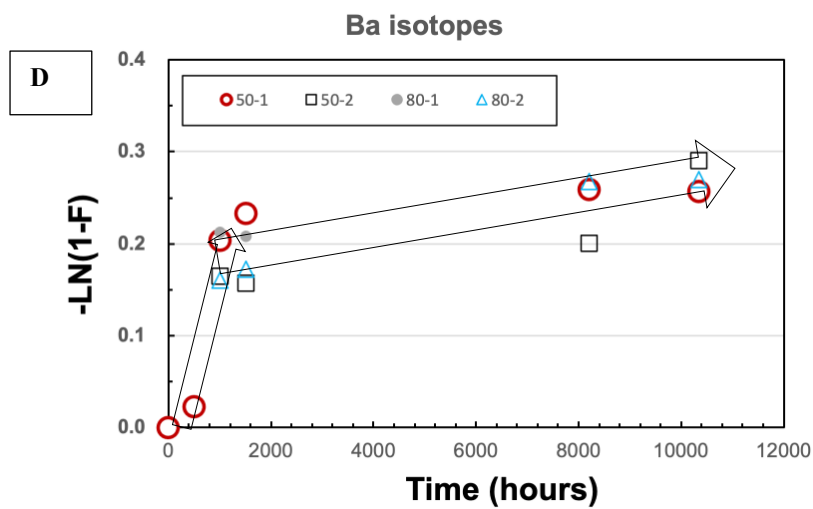

Fig. S3. Relationships between the degrees of isotope exchange with isotope ratios and time. Progresses of isotope exchange toward isotope equilibrium (F) and aqueous sulfur (A) and barium (B) isotopes compositions. $\delta^{34}S$ and $f_{137}$ in the aqueous solutions linearly increased and decreased with F, respectively. Extrapolations back to F=0 at time zero are consistent with the measured initial solution S and Ba isotopes, and extrapolations to F=1 resulted in approximately the solid S and Ba isotope compositions. S and Ba isotope fractionation in this strongly doped system is negligible. (C) and (D) Natural logarithm of fractions of S and Ba isotopes exchanged in aqueous solutions with barite as a function of time. These plots were used to retrieve the bulk isotope exchange rates from Eq. S4.

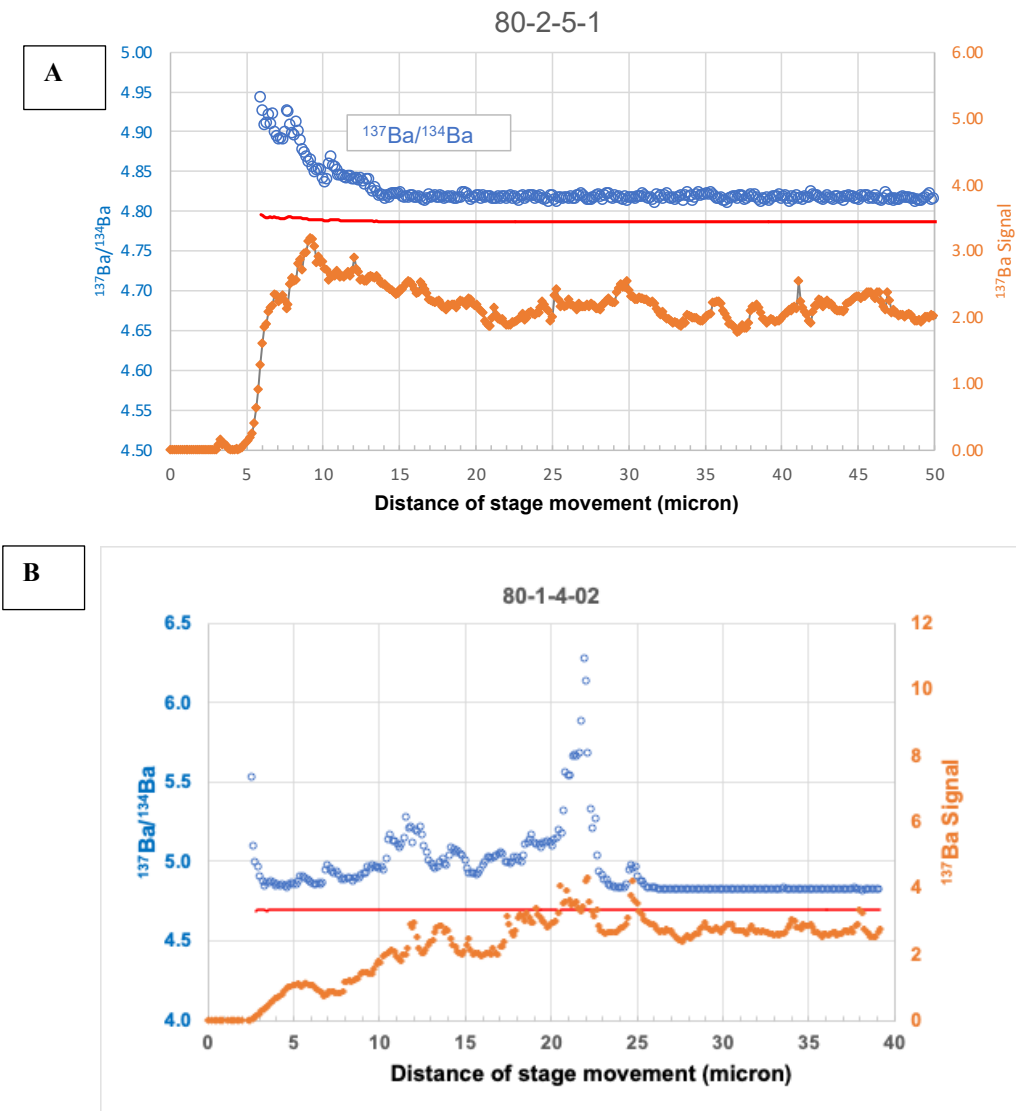

**Fig. S4a,b. Representative barium isotopes LA cross-sections**. Isotopic profiles start from the grain boundary at the left into the barite interior to the right. $^{137}Ba/^{134}Ba$ (blue circles, left axis), $^{137}Ba$ (orange diamonds, right axis), and $^{136}Ba/^{134}Ba$ (red squares, right axis). The detection of $^{137}Ba$ on the left indicated that the beam had started to cover the barite grain. The constant $^{136}Ba/^{134}Ba$ ratios of the undoped isotopes (red squares, as internal control) authenticate the detected $^{137}Ba$ enrichment at the rim and in the middle of barite grains. (**A**) Barite sample (80-2-5) reacted for 10,360 h; and (**B**) Barite sample (80-1-4) reacted for 8,208 h at 80 °C with solutions enriched in $^{137}Ba$ to 29.89% and 55.55%, respectively.

**Table S1. Symbols, notations, and abbreviations in this paper**

| | |
|---|---|
| $m_i$ | concentration of species $i$, mol/kgw |
| $[i]$ | activity of species $i$ |
| $f_{\#}^{p}$ | fraction of isotope # in phase $p$, e.g., $f_{137}^{aq}$ for the fraction of $^{137}$Ba in the aqueous solution |
| $J^{net}$ | Net flux of a species from aqueous solution to the solid, mol m$^{-2}$ s$^{-1}$ |
| $J^+$ | flux of a species from aqueous solution to the solid, mol m$^{-2}$ s$^{-1}$ |
| $J^-$ | flux of a species from the solid to aqueous solution, mol m$^{-2}$ s$^{-1}$ |
| $k$ | rate constant, mol m$^{-2}$ s$^{-1}$ |
| $[Ba^{2+}]/[SO_4^{2-}]$ | aqueous barium-to-sulfate ratio |
| $u_d, u_a, u,$ | The detachment, attachment, and net rate in mol m$^{-2}$ s$^{-1}$, respectively |
| $U_d$ | $U_d = u_d/C_{Ba}^{barite}$. $U_d$ has the unit of m/s, and can be viewed as the linear detachment rate. |
| $w$, or $\Omega$ | Activity product |
| $K$ | Equilibrium constant |
| $K_{sp}$ | Solubility product |
| $SI$ | Saturation index, log(Q/K) |
| $F$ | Degree of isotope exchange |
| $R$ | Gas constant |
| $T$ | Temperature |
| $i$ | Initial |
| $eq$ | Equilibrium |
| $t$ | Time |
| $k$ | Bulk rate constant |
| SEM | Scanning Electron Microscopy |
| NCF | Nanoscale Characterization Facility |
| VCDT | Vienna-Canyon Diablo Troilite |
| BET | Brunauer-Emmett-Teller surface area |
| XRD | X-Ray Diffraction |
| EDS | Energy-Dispersive X-ray Spectroscopy |
| ICP-OES | Inductively Coupled Plasma - Optical Emission Spectrometry |
| IC | Ion Chromatography |
| MC-ICP-MS | Multiple Collector-Inductively Coupled Plasma-Mass Spectrometry |
| SSB | Standard-Sample-Bracketing |
| NIST SRM3104a | National Institute of Standards and Technology (NIST) Ba standard |
| NA | natural abundance |
| SEM | Scanning Electron Microscopy |

Table S2. Matrix of Ba and S isotope exchange experiments

| Expt series | 50-1 | 50-2 | 80-1 | 80-2 |
|---|---|---|---|---|
| Temperature | 50 °C | | 80 °C | |
| Target [Ba] ($\mu$M) | 13.80 | 13.80 | 15.85 | 15.85 |
| Target [SO$_4$] ($\mu$M) | 13.80 | 13.80 | 15.85 | 15.85 |
| BaCl$_2$ stock soln (g) | 0.15 | 0.23 | 0.17 | 0.27 |
| Isotope stock soln (g) | 0.69 | 0.29 | 0.79 | 0.33 |
| Na$_2$SO$_4$ stock soln (g) | 0.19 | 0.19 | 0.22 | 0.22 |
| Ba/SO$_4$ | 1 | 1 | 1 | 1 |
| pH | 5 | 5 | 5 | 5 |
| SI | 0 | 0 | 0 | 0 |
| Solid (g) | 0.0303 | 0.0303 | 0.0303 | 0.0303 |
| liquid (g) | 100 | 100 | 100 | 100 |
| grain size | 18-25 mesh | 18-25 mesh | 18-25 mesh | 18-25 mesh |
| $^{137}$Ba/$^{138}$Ba | 1.4708 | 0.5170 | 1.4842 | 0.5060 |
| $^{136}$Ba/$^{138}$Ba | 0.1100 | 0.1096 | 0.1100 | 0.1096 |
| $f_{137}$ Ba initial | 55.55 | 29.89 | 55.55 | 29.89 |

**Table S3. Ba isotope abundances in the barite reactant**

|         | $^{130}Ba$ | $^{132}Ba$ | $^{134}Ba$ | $^{135}Ba$ | $^{136}Ba$ | $^{137}Ba$ | $^{138}Ba$ |
|---------|--------|--------|-------|-------|-------|--------|--------|
| Natural | 0.1058 | 0.1012 | 2.417 | 6.592 | 7.853 | 11.232 | 71.699 |

Table S4. Experimental isotope exchange rates (×$10^9$ mol m$^{-2}$ s$^{-1}$)

| exp # | hrs | $J_{Ba}$ | $J_S$ | $J_{Ba}$ 22 °C | $J_S$ 22 °C | $J_{Ba}$ 4 °C | refs |
|---|---|---|---|---|---|---|---|
| 22 | 10 | 1.36 | | | | | (*19*) |
| 22M | 0.75-24 | 0.63 | 0.63 | | | | (*67*) |
| 50-1, 50-2 | 0-1008 | 0.843 | 0.843 | 8.35x10$^{-3}$ | 8.35x10$^{-3}$ | | This study |
| 80-1, 80-2 | 0-1008 | 0.896 | 0.959 | 8.35x10$^{-3}$ | 8.35x10$^{-3}$ | | This study |
| 50-1, 50-2 | 1008-10360 | 0.0211 | 0.0211 | 8.35x10$^{-3}$ | 8.35x10$^{-3}$ | | This study |
| 80-1, 80-2 | 1008-10360 | 0.0484 | 0.0484 | 8.35x10$^{-3}$ | 8.35x10$^{-3}$ | 0.00417* | This study |

*Extrapolated to 4 °C at marine sediment conditions from 80 °C using an activation energy of 26.2 kJ mol$^{-1}$. The far-from-equilibrium dissolution rate from Dove and Czank is 0.224 ×$10^9$ mol m$^{-2}$ s$^{-1}$.

Table S5. Fit parameters for $f_{aq}$ in different experimental series

|  | $f_{aq,t=\infty}$ | $f_{aq,t=0}$ | $\tau_2$ (hrs) | $r^2$ | $f_{bs}$ at $t$ |
|---|---|---|---|---|---|
| Series 50-1 | 0.449967 | 0.5555 | 940.36 | 0.907318 |  |
| Series 50-2 | 0.257467 | 0.2989 | 1002.65 | 0.911024 |  |
| Series 80-1 | 0.343586 | 0.5555 | 2388.66 | 0.971541 | 0.2061 (8208 h) 0.2026 (10360 h) 0.2000 (10360 h) |
| Series 80-2 | 0.253760 | 0.2989 | 1121.19 | 0.991630 | 0.2209 (10360 h) |

Table S6. Fit results of $f_{137}$ diffusion profiles in barite assuming $w_0 = 0.811$

| Profile | Time (h) | $f_{137,x=0}$ | $D_{Ba}$ (m²/s) | $u_d$ (mol/m²/s) | $L_{diss}$ (nm) | $L_{dif} = \sqrt{Dt}$ (nm) | $r^2$ |
|---|---|---|---|---|---|---|---|
| 80-2-5 | 10360 | 0.22665 | $6.1 \times 10^{-24}$ | $2.2 \times 10^{-11}$ | 2.4 | 15.1 | 0.99788 |
| 80-1-5a | 10360 | 0.20680 | $6.9 \times 10^{-24}$ | $4.0 \times 10^{-12}$ | 0.43 | 16.0 | 0.9976 |
| 80-1-5b | 10360 | 0.20250 | $8.3 \times 10^{-24}$ | $4.1 \times 10^{-12}$ | 0.45 | 17.6 | 0.9969 |
| 80-1-4 | 8208 | 0.19895 | $1.31 \times 10^{-24}$ | $1.7 \times 10^{-12}$ | 0.18 | 6.2 | 0.9943 |